\documentclass[aps,prl,twocolumn,groupedaddress,showpacs]{revtex4}
\usepackage{graphicx}  

\begin{document}

\title{Spin-Boson Theory for Magnetotransport in Organic Semiconducting Materials}

\author{Yao Yao}\affiliation{Department of Physics and State Key
Laboratory of Surface Physics, Fudan University, Shanghai 200433, China}

\author{Wei Si}\affiliation{Department of Physics and State Key
Laboratory of Surface Physics, Fudan University, Shanghai 200433,
China}

\author{Xiaoyuan Hou}\affiliation{Department of Physics and State Key
Laboratory of Surface Physics, Fudan University, Shanghai 200433,
China}

\author{Chang-Qin Wu}\email[Electronic address: ] {cqw@fudan.edu.cn} \affiliation{Department of Physics and
State Key Laboratory of Surface Physics, Fudan University,
Shanghai 200433, China}

\date{\today}
\begin{abstract}

We present a spin-boson theory for magnetotransport in organic
semiconducting materials, on the basis of a coupling between
charge carriers' spin and a local bosonic environment, which is
shown to be an irreducible ingredient in understanding of the
anomalous organic magnetoresistance (OMR). Among those composing
this environment triplet-excitons play a basic role. The
incoherent hopping rate between molecules is calculated to give
out the fundamental behavior of OMR. The underlying mechanism is
revealed from the calculation of entanglement, represented by the
von Neumann entropy, between the carrier's spin and bosons. We
also obtain the dependence of OMR on the bias voltage, the
spin-boson coupling, and the boson frequency. The results obtained
from the theory are in good agreement with experiments.
\end{abstract}

\pacs{73.43.Qt, 72.20.Ee, 71.35.-y}

\maketitle

The discovery of room-temperature, low-field magnetoresistance
(MR) in organic light-emtting devices (OLEDs) was one of the major
achievements of spintronics in the last decade.\cite{First,Dediu}
Compared to its inorganic counterparts, a sizable organic MR (OMR)
is relatively easy to be obtained, showing extensive potential in
magnetically controlled applications.\cite{Vardeny} Yet, after
years of intense research, the origin of this magnetic field
effect out of these nonmagnetic materials remains controversial.
Traditional theories of MR in inorganic materials, such as that
based on Lorentz force or spin-orbit coupling, are recognized to
be invalid in organic materials.\cite{SOC} The hyperfine
interaction (HFI) caused by hydrogen nuclei in organic molecules
has been thought to be significant,\cite{Hyperfine} which is
verified very recently by an experiment on isotope effect in spin
response of $\pi$-conjugated polymers.\cite{HFIMEL} Meanwhile, a
similar experiment in the Alq$_3$ showed the response is isotope
independent,\cite{Gillin} which seems interactions other than HFI
may dominate the spin process. A bipolaron model\cite{Bipolaron}
using Monte Carlo (MC) simulation gave a mechanism for both of
positive and negative OMR at high density of charge carriers. The
MC method has also been successfully extended to study spin
diffusion in spin valve effect based on incoherent hopping of a
charge carrier and coherent precession of its
spin.\cite{SpinDynamic} Moreover, there are experiments indicating
that the behavior of OMR at low field might be different with that
at relatively high field.\cite{Vardeny,Ours} In a word, a
comprehensive understanding of the experimental phenomena will be
enlightening.

A proper theory that describes the intrinsic physics of OMR should
incorporate the following considerations: (i) A prominent
characteristic of charge transport in organic semiconducting
materials is the incoherent hopping between molecules;\cite{Review1}
(ii) The spin decoherence time is about $0.5\mu$s,\cite{SpinDeco1}
much longer than the time ($\sim$100$n$s in Alq$_3$) for carriers
transporting in devices.\cite{SpinDeco2} Further, the intermolecular
hopping was observed to be spin-conserving,\cite{SpinConserve} which
implies the motion of carriers' spin is coherent;\cite{SpinDynamic}
(iii) The OMR is robust under room temperature.\cite{Room} Even
though the Zeeman splitting energy of about 100$\rm mT$ is much
smaller than the thermal energy, the effect of magnetic field will
not be smeared out by the thermal fluctuation, for the motion of
spins being coherent in the whole process. For the same reason, the
effect is little influenced by all other things that are not
interacting with spins; (iv) There are contradictions among various
experiments on the role of excitons. On one hand, these processes
associated with excitons, such as intersystem crossing between
singlet and triplet excitons\cite{Vardeny,Hu}, and triplet-triplet
annihilation\cite{Xiong,Single} were regarded as the main response
to magnetic field, which implies the role is essential. This might
be complemented by a recent experiment of transient
electroluminescence, which revealed that OMR could be observed only
when the carriers and excitons coexist in the system.\cite{LiFeng}
On the other hand, there are also evidences that the excitons are
irrelevant to the OMR, for example, the measurement of the
intersystem crossing rate shows it is magnetic field
independent.\cite{SpinConserve,Comment} As a result, it can be
realized from above that all mentioned mechanisms overlook the
scattering of carriers from excitons, which should be a {\it
primary} role of excitons in the magnetotransport.

With those considerations, we propose a spin-boson theory for
magnetotransport in this Letter. The Hamiltonian we suggest for
the theory is expressed as
\begin{eqnarray}
H=\sum_n H_n+H^{\prime},\label{Hami}
\end{eqnarray}
where the intramolecular part of the $n$-th molecule
\begin{eqnarray}
H_n&=&\sum_{\alpha}\left[\hbar\omega_{n,\alpha}b^{\dag}_{n,\alpha}
b_{n,\alpha}+\gamma_{n,\alpha}(b^{\dag}_{n,\alpha}+b_{n,\alpha})S^{z}_n
\right]\nonumber\\
&&+~g\mu_B {\bf B}\cdot{\bf S}_n \label{intra}
\end{eqnarray}
describes the spin interaction of a charge carrier with a local
bosonic environment in an external magnetic field ${\bf B}$. Here,
${\bf S}_n$ ($S^z_n$) is the (z-direction) spin operator of a
carrier, $b^{\dag}_{n,\alpha}$ ($b_{n,\alpha}$) creates
(annihilates) a boson that is the $\alpha$-th mode of the
environment, $\omega_{n,\alpha}$ the corresponding frequency of
bosons, $\gamma_{n,\alpha}$ the spin-boson coupling, $g$ the
Land\'{e} factor that is set to be the commonly accepted value
$2.0$, $\mu_B$ the Bohr magneton, and $z$ direction is chosen
along that of the specific bosonic mode at each molecule. As an
important characteristic of organic materials, when a carrier hops
into a molecule, it is immersing in a complex and disordered
surrounding medium. This medium could be treated as an environment
composed of a number of bosons.\cite{Review2} Many factors
contribute to the medium, such as local molecular vibration modes,
hydrogen nuclear spin, and excitons, in a way that depends on
materials.\cite{HFIMEL} Especially, triplet excitons should play a
basic role for their long lifetime and weak diffusive ability
compared with that of charge carriers.\cite{Book} The environment
could be regarded to be an effective magnetic field defined as
\begin{equation}
B_{0n}=\frac{1}{g\mu_B}\sum_\alpha \gamma_{n,\alpha}\langle
b_{n,\alpha}^\dagger+b_{n,\alpha}\rangle\label{B0}
\end{equation}
to the carriers' spin plus quantum fluctuation that will be shown to
be crucial to give the magnetic field effect. The intermolecular
part of (\ref{Hami})
\begin{eqnarray}
H^{\prime}=\sum_{n,n',\sigma}J_{nn'}\left(c^{\dag}_{n,\sigma}c_{n',\sigma}+{\rm
h.c.}\right)
\end{eqnarray}
describes the carrier's hopping between molecules.
$c^{\dag}_{n(n'),\sigma}$ ($c_{n(n'),\sigma}$) creates
(annihilates) a carrier with spin $\sigma$ at $n(n')$-th molecule,
and $J_{nn'}$ the overlap integral of wavefunctions between
molecules. This term includes all the magnetism independent
factors in conventional treatment of organic charge transport,
such as disorders of molecular energies and intermolecular
distances, which ensures that we could only consider a
spin-related Hamiltonian for $H_n$.

Considering the characteristic of organic materials as discussed
above, we could write the density matrix of the system as a direct
sum of all local density matrices of molecules, that is,
\begin{equation}
\rho=\rho_1\oplus\rho_2\oplus\cdots\rho_n\oplus\cdots,\label{rho}
\end{equation}
and divide the whole process into two steps, intramolecular
evolution and intermolecular hopping, to calculate the incoherent
hopping rate of a single charge carrier. The dynamical evolution
follows the equation of motion,
\begin{eqnarray}
i\hbar\frac{\partial}{\partial
t}\rho_{n}=[H_n,\rho_{n}].\label{Evo}
\end{eqnarray}
$H_n$ could be numerically diagonalized by constructing the
following basis, indexed by a set of integer numbers
$\{l_\alpha\}$,\cite{Wu}
\begin{eqnarray}
|\{l_\alpha\},s\rangle=\prod_\alpha\frac{e^{-u^2_\alpha/2}}{\sqrt{l_\alpha!}}
(\hat{b}_\alpha^{\dag}+su_\alpha)^{l_\alpha}e^{-su_\alpha\hat{b}_\alpha^{\dag}}|\{0\},s\rangle,\label{boson}
\end{eqnarray}
where $u_\alpha=\gamma_\alpha/\hbar\omega_\alpha$ is the
displacement of bosons, $s=+1$ for spin up and $-1$ for spin down,
and for simplicity we have dropped the molecular index $n$. Here,
$u_\alpha$, depending on the spin-boson coupling, is approximately
equal to the Huang-Rhys factor $S$ ($=\lambda/\hbar\omega$ with
$\lambda$ is the reorganization energy), which is commonly of the
order $\sim 1$ in organic materials.\cite{HuangRhys} Throughout
this work, we set the cutoff number of bosons to be $80$ to ensure
the calculation is convergent.

The second step is the \textit{incoherent} hopping between
molecules, whose rate could be derived in terms of the Fermi
golden rule,\cite{qm}
\begin{eqnarray}
\nu\approx\frac{\tau_{\rm if}}{\hbar^2}|\langle
f|H^{\prime}|i\rangle|^2\equiv\frac{\tau_{\rm if}}{\hbar^2}{\rm
Tr}\left\{H^{\prime}\rho_fH^{\prime}\rho_i\right\},\label{a0}
\end{eqnarray}
where $\tau_{\rm if}\equiv
4\hbar^2\sin^2[(E_f-E_i)t_d/2\hbar]/(E_f-E_i)^2t_d$ with $t_d$
being the decoherence time within which the hopping is coherent,
and $\rho_i$ ($|i\rangle$) and $\rho_f$ ($|f\rangle$) are the
initial and final density matrix (state) with the energy $E_i$ and
$E_f$, respectively, expressed in Eq.~(\ref{rho}). $t_d$ is a
quantity determined approximately by the molecular structure
relaxation, which gives it's at the order of 0.1ps.\cite{Review1}
It's noted that this treatment is qualitatively similar to the
Franck-Condon principle,\cite{Review2} which has been widely used
in organic electronics as well as the spin-dependent exciton
formation.\cite{FCinEG} To determine the initial and final state
in (\ref{a0}), we consider the following process. At a time when
the process begins, a carrier with some spin ${\bf s}$ hops onto
the $n$-th molecule, and the density matrix could be expressed as
a direct product
\begin{eqnarray}
\rho_n(t=0)=\varrho_{n}({\bf s})\otimes\tilde{\varrho}_{n},
\end{eqnarray}
where $\varrho$ denotes the ($2\times2$) spin density matrix of a
carrier while $\tilde{\varrho}$ for bosons. Right at this moment,
the interaction between the carrier and local bosonic modes is
switched on, and then they become \textit{entangled}. The system
evolutes following Eq. (\ref{Evo}) till the carrier hops to the
unoccupied $n'$-th molecule at time $t_w$, called waiting time,
whose average value could be estimated by the mobility of carriers
and bias voltage. For organic materials, $t_w\gg t_d$. After the
hop, the carrier's spin and bosons on the $n$-th molecule
disentangle. The spin state keeps unchanged because of
spin-conserving, while the bosons will reorganize. An adiabatic
elimination procedure technique is used for the spin
state,\cite{Entangle} which gives the spin density matrix at the
$n'$-th molecule as
\begin{equation}
\varrho_{n'}({\bf s})={\rm Tr}_b\{\rho_n(t_w)\},
\end{equation}
where the trace with subscript $b$ is that over all bosonic
degrees of freedom. For the reorganization of bosons, it depends
on the relative ratio of the relaxation time of the bosonic
environment ($t_r$), which is not explicitly known, and that of
molecular structure ($\sim t_d$). In case $t_r \gg t_d$, the
relaxation is negligible, so the bosonic density matrix becomes
\begin{equation}
\tilde{\varrho}_{n}={\rm Tr}_s\{\rho_n(t_w)\},
\end{equation}
where the trace with subscript $s$ is that over spin degree of
freedom. But for $t_r \ll t_d$, the relaxation is complete, and
$\tilde{\varrho}_n$ should return to its equilibrium state, in
which all $u_\alpha=0$. For an arbitrary ratio of $t_r/t_d$, we
could introduce phenomenological parameters $\eta_\alpha$
depending on $\omega_\alpha$ between 0 and 1. Thus,
$\tilde{\varrho}_n$ in Eq.~(\ref{boson}) should be that with all
$u_\alpha$ replaced by $u_\alpha^{*}(\equiv \eta_\alpha
u_\alpha)$. It will be seen later that the result is insensitive
to the choice of $\eta_\alpha$, \textit{i.e.}, the relaxation of
bosonic environments.

Now, the hopping rate could be derived directly from
Eq.~(\ref{a0}) as
\begin{equation}
\nu_{n\rightarrow n'}=\frac{\tau_{nn'} J^2_{nn'}}{\hbar^2}{\rm
Tr}\{\tilde{\varrho}_{n}\varrho_{n'}({\bf s})\rho_n(t_w)\},
\end{equation}
since it concerns only two molecules. Considering the random
directions of a carrier's initial spin and its coupling with
bosons, the hopping rate we calculate below will be averaged over
all directions of the carrier's spin and that of applied magnetic
field since the spin-boson coupling has been fixed along the
$z$-direction.

\begin{figure}
\includegraphics[angle=0,scale=0.85]{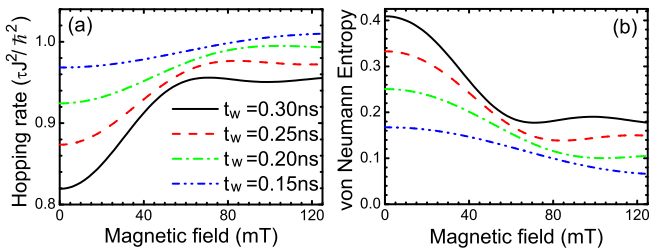}
\caption{Magnetic-field dependence of (a) the hopping rate $\nu$
(in unit of $\tau J^2/\hbar^2$) between two molecules and (b) the
von Neumann entropy for various waiting times $t_w$.
$\gamma=1.0\mu$eV and $\hbar\omega=2.0\mu$eV.}\label{rate}
\end{figure}

Within the framework of perturbation theory, the interaction between
different bosonic modes is not significant, so it is meaningful to
investigate the system with an environment of a single bosonic mode.
Hereafter we omit those molecular and boson's indexes if it would
not create confusion. At first, the calculated hopping rate is shown
in Fig.~\ref{rate}(a),
from which we can see its strong dependence on external magnetic
field as expected. We have set $\gamma=1.0\mu$eV and
$\hbar\omega=2.0\mu$eV, which correspond to an effective magnetic
field $B_0 (\equiv 2\gamma^2/\hbar\omega g\mu_B)=8.6$mT evaluated
from Eq.~(3). This field strength is similar with that used in HFI
model for OMR\cite{Hyperfine} and extracted from muon
experiment\cite{Muon}. Accordingly, a saturation is reached for
the magnetic field being $50$ - $100$mT depending on the waiting
time $t_w$. It is clearly seen that the magnetic field dependence
becomes stronger with a larger $t_w$. The slow waviness in those
curves is due to the Rabi oscillation between spin states.

The observed magnetic field dependence originates from the
incoherent hopping, which disentangle the spin and bosonic
environment. To show it we calculated the von Neumann
entropy\cite{Book}, which is defined as ${\rm
Tr}(\tilde{\varrho}\log_2\tilde{\varrho})$. The result in
Fig.~\ref{rate}(b) implies that the entanglement weakens the
ability of a carrier hopping out of the environment. The external
magnetic field changes the entanglement between the carrier and
the bosonic environment, and then the magnetic field effect
arises. This is the basic mechanism of OMR from our theory.

\begin{figure}
\includegraphics[angle=0,scale=0.93]{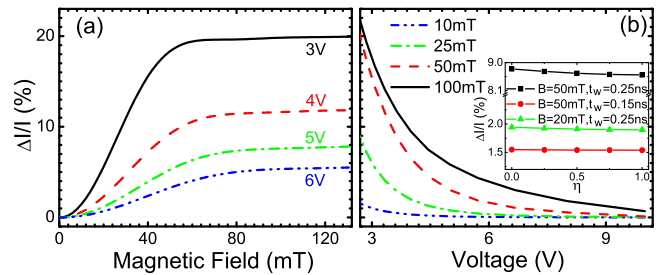}
\caption{(a) Magnetic field dependence of OMR under various bias
voltages. (b) The dependence of OMR on the voltage under various
magnetic fields. Inset shows the OMR is insensitive to
$\eta$.}\label{Field}
\end{figure}

With the hopping rate between molecules in Eq.~(12), we have the
OMR as
\begin{eqnarray}
\frac{\Delta I}{I}\equiv\frac{I(B)-I(0)}{I(0)}=\frac{\langle
\nu(B)\rangle-\langle \nu(0)\rangle}{\langle \nu(0)\rangle},
\end{eqnarray}
where $I(B)$ is the current depending on the magnetic field $B$, and
$\langle\cdot\rangle$ is the average over samplings with a
distribution of $t_w$ here.\cite{disorder} The average of $t_w$
should be proportional to the inverse of bias voltage applied to a
given device, which is expressed as $\langle t_w\rangle=L_D l/\mu
V_b\xi$, with $L_D$ being the thickness of device, $l$ the
intermolecular distance, $\mu$ the mobility of carriers, $V_b$ the
bias voltage, and $\xi$ a factor introduced to account for the
carrier's random hop. For example, we consider a typical OLED
structure ITO (indium tin
oxide)/NPB($N,N^{\prime}$-di-1-naphthyl-$N,N^{\prime}$-diphenylbenzidine)($50$
nm)/Alq$_3$ (tri-(8-hydroxyquinoline)-aluminum)($50$nm)/Al
(aluminum). The effective mobility of electrons in Alq$_3$ is taken
as $\mu\approx 10^{-4}$cm$^2/$Vs, the average distance between
molecules $l\approx 0.5$nm, and $\xi\approx 5$, so we have,
\textit{e.g.}, $\langle t_w\rangle\approx 0.25$ns for $V_b=4.0$V.
Fig.~\ref{Field}(a)
shows the OMR under different bias voltages. The basic line shape
obtained here is in good agreement with a consensus based on
experiments.\cite{First,Hyperfine,LiFeng,Bipolar} Furthermore, we
show the dependence of OMR on bias voltage under different
magnetic field in Fig.~\ref{Field}(b). A decay following the
voltage increase is found, which matches experimental
observations.\cite{LiFeng} For a small bias voltage,
\textit{e.g.}, $V_b<2.0$V, the current is extremely small and the
scattering between excitons becomes important, which is not
included in this work, that might be the origin of positive OMR
observed in experiments.\cite{Bipolar} The $\eta$ dependence of
OMR is given in the inset, which verifies the choice of $\eta$ has
little influence.

\begin{figure}
\includegraphics[angle=0,scale=0.88]{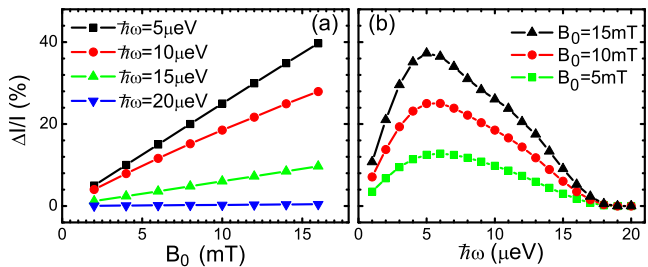}
\caption{OMR dependence on (a) the effective magnetic field $B_0$
and (b) the boson energy $\hbar\omega$ with
$t_w=0.25$ns.}\label{omega}
\end{figure}

Now we come to see the influence of a bosonic mode on the OMR.
Fig.~\ref{omega}(a)
shows the dependence of saturated $\Delta I/I$ on the spin-boson
coupling $\gamma$, which is equivalent to the effective magnetic
field $B_0$, at $t_w=0.25$ns. Obviously, OMR is found to be nearly
proportional to $B_0$ in the range we calculated.
Fig.~\ref{omega}(b) shows the frequency dependence of OMR. As
expected, OMR approaches to vanishing when the frequency becomes
large enough, where the bosonic mode is equivalent to an effective
magnetic field, that is, the (classical) hyperfine
field.\cite{Hyperfine,Ours} From the perspective of entanglement,
it's easy to understand the vanishing OMR since a classical field
does not entangle with the carriers' spin. This also shows the
quantum nature of local bosonic environments is \textit{irreducible}
for OMR.\cite{comment2} It appears that the OMR peaks all reside at
$\hbar\omega\approx 5.0\mu$eV, which is caused by setting the
waiting time at a fixed value $t_w=0.25$ns. Such phenomenon implies
that bosonic modes of compatible energy with the waiting time
contribute more to OMR.

Until now, we have discussed the environment of a single bosonic
mode. In reality, a number of modes should contribute to OMR, and
then the effective magnetic field $B_0$ we used in the one-mode
case could be regarded as that defined in Eq.~(3) within the
perturbation theory. As has discussed, triplet-excitons play a
basic role in the environment, so $B_0$ should be strongly
dependent on the density of excitons that could be adjusted
experimentally in many ways. Moreover, another important physical
quantity to OMR is the boson energy $\hbar\omega$, which should
depend on the constitution of the environment, related to
characters of specific material, such as the mass of atoms and/or
radicals in the molecule, electronic excitation energies. Due to
the nonmonotonicity of OMR on $\hbar\omega$ as shown in
Fig.~\ref{omega}(b), the discrepancy of isotope effect in
different materials\cite{HFIMEL,Gillin} might be understood within
the theory. It is also shown\cite{Xiong} experimentally that the
OMR is strongly associated with the excitation energy by changing
dopants.

In summary, we have proposed a theory based on the coupling
between carriers' spin and a local bosonic environment. The
entanglement is calculated to reveal the underlying mechanism of
the anomalous OMR. The quantum nature of the environment is shown
to be irreducible. The theory establishes the basic line shape of
OMR and its dependence on bias voltage in consistent with a
consensus based on experiments. Triplet excitons are believed to
contribute mostly to this environment, and relevant experiments
are understood in accordance with the theory.

\begin{acknowledgments}
One of authors (C.Q.W.) is grateful to Professors D.-H. Lee, X.
Sun, X.C. Xie, and L. Yu for helpful discussions. The work was
supported by the NSF of China, the National Basic Research Program
of China (Grant No. 2009CB929204), and the EC Project OFSPIN
(Grant No. NMP3-CT-2006-033370).
\end{acknowledgments}


\begin{thebibliography}{99}
\bibitem{disorder} In principle, we should take the average over all kinds of
disorder in organic materials, which include that of molecular
energy, intermolecular distance, and boson modes with different
coupling strengths and frequencies, and the result will be changed
only quantitatively.
\bibitem{comment2} This conclusion does not conflict with the fact the
(classical) hyperfine field has been shown to play a crucial role
in exciton formation.\cite{Ours}
\end{thebibliography}
\end{document}